# Observation of optical spatial solitons in a highly nonlocal medium

Claudio Conti, Marco Peccianti, and Gaetano Assanto

N*OO*EL - Nonlinear Optics and OptoElectronics Laboratory

National Institute for the Physics of the Matter (INFM), University "Roma Tre"

Via della Vasca Navale 84, 00146 Rome, Italy

We report on the observation and quantitative assessment of self-trapped pulsating beams in a highly non-local nonlinear regime. The experiments were conducted in nematic liquid crystals and allow a meaningful comparison with the prediction of a scalar theory in the perturbative limit, while addressing the need for beyond-paraxial analytical treatments.

Several remarkable nonlinear physical phenomena may be classified as solitons, or solitary waves. Among the most famous is "the Great [water] Wave of Translation", originally reported by J. S. Russell and considered the first documented observation of solitons.[1] A fundamental characteristic unifies soliton phenomena: the discreteness arising in the framework of continuous fields (such as in electro- and fluid- dynamics). From a mathematical viewpoint, solitons are solutions of specific differential equations which are nonlinear and *integrable*. In general, however, nonlinear waves which keep a distinctive and unchanged identity along their propagation may be considered solitons, or solitary waves (SW), even when they are described by non-integrable systems.[2] In the applied nonlinear optics community such a distinction is often set aside, [3] as it will in this work.

Several approaches have been available to model SW: from heavy computations to sophisticated analytic transforms. More recently, a simple theoretical model was advanced which describes complex soliton-like dynamics and has direct relevance to the novel beam propagation we present here. In particular, Snyder and Mitchell described the propagation of optical spatial solitary waves (i. e., non-diffracting light beams which self-trap owing to the nonlinearity) in a highly non-local nonlinear medium (HNNM), i. e. a medium (described by an integro-differential constitutive relationship in the spatial coordinates) in which the optical field modifies the index of refraction in a *kernel* region much larger than the beam waist [4]. The model in Ref. [4], however, was attributed a limited practical relevance, mainly because of the apparent lack of a physical medium (i. e., an HNNM) able to support such solitons. Nevertheless, nonlocal media with a nonlinear response have been experimentally investigated to outline their specific features in relation to light localization and optical solitons, and among them atomic gases, [5] photorefractive and liquid crystals (LC).[6-12] Photorefractives are highly anisotropic crystals with a finite degree of non-locality due to carrier drift and diffusion [6-9], while LC are liquids with a significant degree of molecular order and birefringence under appropriate anchoring. [10-12]

In recent years, spatial optical solitons have been observed in a variety of nonlinear media and have attracted considerable interest for their potential applications to signal processing and communications.[3,13-14] Such observations were mostly carried out in materials effectively described in terms of a local response (where the relation between the field and the induced nonlinear polarization can be taken as punctual), i. e., cubic or quadratic or photorefractive [15].

In this Letter, after showing theoretically that a highly-nonlocal response is appropriate to describe light propagation in nematic LC, in such HNNM we experimentally investigate nonlocal optical solitons and carry out a quantitative comparison with the model. Their features strongly differ from those of self-trapped beams in local media, and we trust that the results presented hereby are relevant in all those fields where non-locality cannot be neglected and a local picture fails; among them, Bose-Einstein condensation [16] and plasmas [17] are worth mentioning. Not only understanding the interplay between non-locality and nonlinearity is fundamentally important, but it also leads to intriguing applications of spatial solitons: for instance, the recently demonstrated non-trivial logic gates based on non-local interactions between optical SW [18].

**Theory.** With reference to the highly-nonlocal regime, while the theory developed in Ref. [4] provides insight and general guidelines on the dynamics of spatial solitons in HNNM, a more specialized description is required to model the response of a particular medium such as a nematic LC.

The experimental geometry of our nematic liquid crystal cell for planar orientation is sketched in Fig. 1. With proper anchoring boundaries, the molecular director $\theta(x,y,z)$ distribution can be initially adjusted to a background value by an externally-applied low-frequency voltage.[19] By setting $\theta_0(x,y,z) = \theta_0 \cong \pi/4$ with respect to the linear spatial polarization of an incoming light beam, in the paraxial approximation using standard expressions for LC [10] at the first-order, from

$$4K\nabla_\perp^2 \theta + \varepsilon_0 \Delta\varepsilon \sin(2\theta)|E|^2 = 0 \qquad (1)$$

for finite radial distances from the beam axis (in $r = 0$), it stems that the dipole-induced perturbation is much wider that the beam itself, with:

$$\theta \cong \theta_0 + \Delta\theta - \frac{\varepsilon_0 n_a^2}{16K}|E(r=0,z)|^2 r^2 \qquad (2)$$

where $\Delta\theta$ is a small ($\Delta\theta << \theta_0$) optically-induced correction to the background value $\theta_0$ of the director angle, $r$ the cylindrical radial coordinate, $\varepsilon_0$ the vacuum permittivity and $E$ the electric field amplitude in the beam. $K$ is the LC elastic constant (conveniently taken equal for splay, bend and twist of the molecules) and $n_a^2 = \Delta\varepsilon = n_\parallel^2 - n_\perp^2$ the dielectric anisotropy, i. e., the difference between the permittivities along the two principal axes of the LC ellipsoidal molecule (for the liquid crystal E7, $K = 1.5 \times 10^{-11}$ N, $n_a \cong 0.6$). The director distribution Eq. (2) relates to a likely HNNM, although it should not be extended to large $r$, where it clearly becomes unphysical. From

$$2ik\partial_z E + \nabla_\perp^2 E + \frac{w^2}{c^2} n^2(q) E - k^2 E = 0, \tag{3}$$

which describes the propagation of a linearly x-polarized light wave in a perturbed director field ($n(\theta) = n_\parallel n_\perp / \sqrt{n_\parallel^2 \cos^2(\theta) + n_\perp^2 \sin^2(\theta)}$), about $\theta_0 = \pi/4$ the Foch-Leontovich equation for the beam-envelope in the limit of a thick cell reads

$$2ik\frac{\partial E}{\partial z} + \Delta E - \frac{p^2 e_0 n_a^4}{4l^2 K} |E(r=0,z)|^2 r^2 E = 0 \tag{4}$$

being $k = 2\pi n(\theta_0)/\lambda$ the propagation constant, $\lambda$ the wavelength (we take $\lambda = 1.064 \mu m$) and $n(\theta_0)$ the background refractive index ($n(\pi/4) \cong 1.6$). Eq. (4) describes the re-orientational nonlinearity of LC, and supports a Gaussian solution with a waist $W(z)$ (in intensity) periodically varying along the propagation distance. While coupled equations (1) and (3) can be used for numerical analysis of the phenomenon, (4) is a rough model indicating the trends of measurable quantities in our experiments. Its solution can be found by writing $E = E_0 \exp(iQ(z) + ikr^2/2q(z))$. Being $|s(z)|^2 = (E_0/|E|)^2$, $q = s/s'$ and $Q' = 1/q$, the complex parameter $s(z)$ obeys

$$\frac{d^2 s}{dz^2} + \frac{\varepsilon_0 n_a^4 E_0^2}{16 n^2 K} \frac{s}{|s|^2} = 0, \tag{5}$$

which, for a flat-phase gaussian input, can be reduced to the equation of a normalized nonlinear one-dimensional oscillator, with position $\rho = |s|^2 \sqrt{P/P_S}$ in a parabolic-like potential: $\ddot{\rho} = -\partial U/\partial \rho$, $U = \ln(\rho) + 1/2\rho^2$ (dots denote derivatives with respect to a normalized propagation coordinate), with $r(0) = \sqrt{P/P_S}$ and $\dot{r}(0) = 0$.

An alternative insightful (albeit less rigorous) way to solve Eq. (4) consists in looking for transverse-localized waves, in the hypothesis that the optical beam is nearly trapped, i. e., $|E(r=0,z)|^2$ is almost z-independent. In essence, the on-axis intensity dependence is reduced to a power dependent term (for a Gaussian solution this is self-consistent), as in Ref. [4]. This approximation to the full dynamics encompassed by coupled Eqs. (1) and (3) accounts for the specific material constants and provides beam evolution laws in analogy to those in Ref. [4]:

$$\frac{W^2(z)}{W_0^2} = 1 + \gamma \sin(\frac{2\pi}{\Lambda} z)^2$$

$$\gamma = \frac{P_s}{P} - 1 \quad ; \Lambda = \frac{8\pi^{3/2}}{\sqrt{2}} \frac{n^{3/2}}{n_a^2} \sqrt{\frac{cK}{P}} W_0$$

(6)

Hence W(z) oscillates with period $\Lambda$, and $W_0$ an initial reference value. P is the optical power, c the light velocity in vacuum, and $P_S = 2\lambda^2 K c n /(\pi n_a^4 W_0^2)$ the *soliton power*.

Thus, the beam generally undergoes nonlinear oscillations, their entity determined by the ratio between the input and the soliton power. The latter fixes the initial position of the particle in the bell-shaped potential. When $P = P_s$ the particle is initially settled in the stationary point, hence it remains still with z, corresponding to an exact soliton excitation. Our treatment above corresponds to assume $|\gamma| \cong 1$, hence it holds for excitations around the soliton value ($P \cong P_S$) and enables to deal with explicit expressions.

Eq. (6), in qualitative agreement with Ref. [4], will be used to interpret the experiments. The dimensionless constant $\gamma$ (i. e., the *harmonic content* of the soliton) identifies various self-trapping regimes. For $\gamma > 0$ the beam diffracts at first before it starts to periodically pulsate, while for $\gamma < 0$ it initially self-focuses. $\Lambda$ grows linearly with $W_0$ and inversely with the square root of the power, because self-trapping stems from a balance between diffraction and self-focusing: the natural tendency of a beam to spread, stronger when $W_0$ is smaller, is overcome by the power-dependent response. In the case $\gamma = 0$ (i. e., $P = P_s$) nonlinearity exactly counteracts diffraction and the beam travels un-modified along $z$. Therefore, (for P  $P_s$) we can state that *spatial solitons in highly non-local media are pulsating beams*.

The previous result can also be interpreted by observing that, according to Eq. (4), the LC system behaves as a graded-index (GRIN) lens with pitch depending not on beam waist, but on beam peak-intensity (or power in the limit of Eq. (6)). This agrees with the definition of an HNNM: the field intensity in a point modifies the medium in a region much larger that the beam spot. When the spot size is very small, the beam does not sense the GRIN and widens, until its waist gets large enough for the lens to become effective. Then, it starts focusing, and the process repeats itself in a cyclic fashion. Noteworthy, despite the specific material setting (LC exhibiting a reorientational response which is far from saturation) and within the paraxial frame, the general SW trend in an HNNM qualitatively resembles the case of a (logarithmic) saturable nonlinearity. [20]

The non-local nonlinearity is able to trap intense and narrow-waist solitary beams. By looking at Eq. (6), we note that when $\gamma > 0$ ($\gamma < 0$) the maximum (minimum) of $W^2$ depends only on material parameters and the power:

$$W_M^2 = \frac{2l^2 Kcn}{\pi n_a^4 P} \quad , \qquad (7)$$

i.e., there is no lower (upper) bound to its minimum (maximum) $W_0$. Although derived within the limit of validity of Eq. (6), the oscillating behavior is the main difference between our HNNM and other responses (such as photorefractive, saturable Kerr, or quadratic) in systems where the SW-waist is a constant determined by the light intensity (left aside the internal modes of a soliton). Noteworthy, Eq. (7) can be recast as

$$PW_M^2 = P_s W_0^2 \qquad (8)$$

with the product $P_s W_0^2 = 2l^2 Kcn/\pi n_a^4$ depending on material parameters and wavelength. Eq. (8) is the existence curve for stationary nonlocal solitons.

**Experimental results.** Fig. 2 shows the measured waist versus $z$, as obtained by analyzing snapshots of the light scattered above the LC cell when SW at $\lambda = 1.064 \mu m$ were launched. Starting at low powers ($\gamma > 0$), the oscillations progressively reduce in amplitude and their period decreases as the excitation goes up. This is in good qualitative agreement with the prediction in Eq. (6), but the experimental accuracy is inherently limited by the acquisition of scattered photons from the SW. To account for this, we modeled our imaging system as a blur, convolving the retrieved intensity profiles with a Gaussian kernel (G). The kernel width was evaluated from the linear diffraction of a known beam, resulting in $G(y) = \exp[-y^2/(4\mu m)^2]$. Due to the blur, an actual $5\mu m$ waist results overestimated by about 50%, but the error is larger than 100% for waists $<1\mu m$. Conversely, if the waist is $>10\mu m$, the error is under 10% (see Fig.1, lower right). Henceforth, we carried out the quantitative comparison with the theory considering only the maximum waist (from Eq.

(7)) and the oscillation period (from Eq. (6)), the latter taken as twice the separation between the first two extrema in the graph of waist versus $z$ (Fig. 2).

Fig. 3a displays the measured maximum SW waist versus power. The best fit is obtained from Eq. (7) by introducing as a parameter the coupling efficiency $\alpha$ of the laser power $P_{in}$ into the soliton-trapped power $P$ (i. e., $\alpha = P/P_{in}$). By minimizing the standard deviation, we found $\alpha = 7\%$. To further the comparison, Fig. 3b shows a plot of $1/\max(W^2)$ versus $P$. The overall linearity is satisfactory and the standard deviation from the model is 4%. Fig. 3c graphs period versus power, and the best fit from Eq. (6). Given the previously evaluated $\alpha$, here the fitting parameter is the minimum waist, yielding $W_0 \approx 1\mu m$. While the latter could not be determined *a priori*, because light enters the sample through the interface and an LC transition layer, such wavelength-size estimate stems from a paraxial model and should, therefore, be taken as the indication of its inherent limit. Finally, in Fig. 3d the quantity $1/\Lambda^2$ is plotted versus $P$. Once more, there is good agreement with the calculation, and the standard deviation is 7%. The overall comparison between data and model is marginally affected by the limited experimental accuracy; however, it emerges clearly that we observed *self-trapped optical beams in a HNNM.* Finally, it is well worth underlining that an additional feature of spatial solitons in HNNM is their collisional behavior, as predicted in Refs. [4] and [21]. Indeed, the recently demonstrated incoherent-like and attractive interactions between SW in the same LC provide an independent proof of their highly-nonlocal dynamics. [22]

A corollary of our model Eq. (6) is that *the minimum waist of the oscillating SW directly relates to its maximum waist and period.* However, since such minimum is estimated comparable to the light wavelength in the medium ($\lambda/n \cong 0.68 \mu m$) and the experimental data are masked by scattering, based on the paraxial approach (Eqs. (4)-(5)) and the small oscillation approximation (Eq. (6)) above we can only state that $W_0$ is rather small. While a vectorial treatment (e.g. Ref. [23]) is needed when dealing with a non-paraxiality figure $C = \lambda/(nW_0) = 4\pi^2 W_M / \Lambda$ close to unity, to circumvent the limits of our linearized analysis leading to Eq. (6) we resorted to a numerical study of light propagation while solving the basic eqs. (1) and (3) for the nonlinear response of our voltage-biased LC cell. The study, carried out with a split-step beam propagator in the paraxial regime, confirms that the SW undergoes oscillations in waist and intensity (up to one order of magnitude), with minimum values of the order of $1 \mu m$. The numerical results ratify the widely oscillating character of the SW as observed in LC, and prompt for a non-paraxial vectorial theory. The latter will be considered elsewhere.

**Conclusions.** Our investigation of an HNNM, employing the nematic liquid crystal E7 in the near infrared, demonstrates that highly nonlocal solitons can be excited and propagate as pulsating beams. In spite of the inherent limits of a paraxial analysis in the small-oscillation approximation, the model and the comparison with data indicate that the beam dynamics is primarily determined by the ratio between input and *soliton* power, with propagation of narrow-waist SW. The inferred non-paraxiality C, in fact, is more than one order of magnitude larger than for self-trapped beams in photorefractive and frequency-doubling crystals. In conclusion, not only nematic LC can be regarded as HNNM and sustain highly non local solitary waves, but they also prompt for the development of non-paraxial nonlinear models in optical soliton propagation.

**Aknowledgements:** we thank C. Umeton (LICRYL, Univ. Calabria, Italy) for the high quality samples, A. Fratalocchi (NOOEL) for some of the numerics, I. C. Khoo (Penn State Univ.), A. Snyder (ANU) and M. Segev (Technion Univ., Israel) for enlightening discussions.

**References and notes**

1. J. S. Russell, Report 14$^{th}$ Meet. British Association for the Advancement of Science (Murray, London), 311 (1844)
2. P. G. Drazin and R. S. Johnson, *Solitons: an introduction*, Cambridge Univ. Press, Cambridge (1989)
3. G. I. Stegeman, D. N. Christodoulides, and M. Segev, IEEE J. Sel. Top. Quantum Electron. **6**, 1419 (2000)
4. A. W. Snyder and D. J. Mitchell, Science **276**, 1538 (1997)
5. D. Suter and T. Blasberg, Phys. Rev. A **48**, 4583 (1993)
6. M. Segev *et al.*, Phys. Rev. Lett. **68**, 923 (1992); G. Duree *et al.*, Phys. Rev. Lett. **71**, 533 (1993); M. Shih *et al.*, Electron. Lett. **31**, 826 (1995)
7. A. V. Mamaev *et al.*, Phys. Rev. A **56**, R1110 (1997)
8. E. DelRe, A. Ciattoni, and A. J. Agranat, Opt. Lett. **26**, 908 (2001)
9. D. Neshev *et al.*, Opt. Lett. **26**, 1185 (2001)
10. N. V. Tabiryan, A. V. Sukhov, and B. Ya. Zel'dovich, Mol. Cryst. Liq. Cryst. **136,** 1 (1986); I. C. Khoo, *Liquid Crystals: Physical Properties and Nonlinear Optical Phenomena*, Wiley, New York (1995)
11. I. C. Khoo, T. H. Liu, and P. Y. Yan, J. Opt. Soc. Am. B **4**, 115 (1987)

12. Apart from the reorientational response of interest here, the thermal nonlocality of dye doped-LC was assessed by J. F. Henninot *et al.* in Mol. Cryst. Liq. Cryst. **375**, 631 (2002)
13. A. D. Boardman and A. P. Sukhorukov, Eds., *Soliton Driven Photonics*, Kluwer Acad. Publ., London (2001); S. Trillo and W. Torruellas, Eds., *Spatial Solitons*, Springer, New York (2001); Y. Kivshar and G. Agrawal, *Optical solitons: from fibers to photonic crystals*, Acad. Press, London (2003)
14. G. I. Stegeman and M. Segev, Science **286**, 1518 (1999)
15. L. Torner and G. I. Stegeman, Opt. Photon. News **12**, 36 (2001)
16. L. Khaykovich *et al.*, Science **296**, 1290 (2002); K. E. Strecker *et al.*, Nature **417**, 150 (2002)
17. V. I. Kolobov and V. A. Godyak, IEEE Trans. Plasma Sci. **26**, 955 (1998)
18. M. Peccianti *et al.*, Appl. Phys. Lett. **81**, 3335 (2002)
19. M. Peccianti and G. Assanto, Phys. Rev. E **65**, R035603 (2002)
20. A. W. Snyder and D. J. Mitchell, Opt. Lett. **22**, 16 (1997)
21. D. J. Mitchell and A. W. Snyder, J. Opt. Soc. Am. B **16**, 236 (1999)
22. M. Peccianti, K. A. Brzd¹kiewicz, and G. Assanto, Opt. Lett. **27**, 1460 (2002)
23. A. Ciattoni *et al.*, Opt. Lett. **27**, 734 (2002)

**Figure captions**

Figure 1. (**A**) Experimental setup: a 75μm thick (along x) planarly aligned nematic LC cell is voltage biased via thin-film electrodes. A light beam impinges from the left, and propagates along z. A CCD-camera with a microscope monitors the beam evolution along y and z by acquiring images of the scattered light from the top. (**B**) the photo of a typical pulsating soliton. (**C**): plot of relative error e=(measured – actual)/(actual) in evaluating the beam waist from the scattered light, as modeled by a Gaussian blur. The cell thickness is much larger than the beam, justifying the radial symmetry assumption.

Figure 2. Measured SW waist versus propagation distance for various excitation powers $P_{in}$.

Figure 3. Observation of accessible solitons: (**A**) maximum waist versus power; (**B**) inverse square of waist vs. power; (**C**) period vs. power (error bars are negligibly small); (**D**) inverse square of waist vs. period. The solid lines are best fits from the theory.

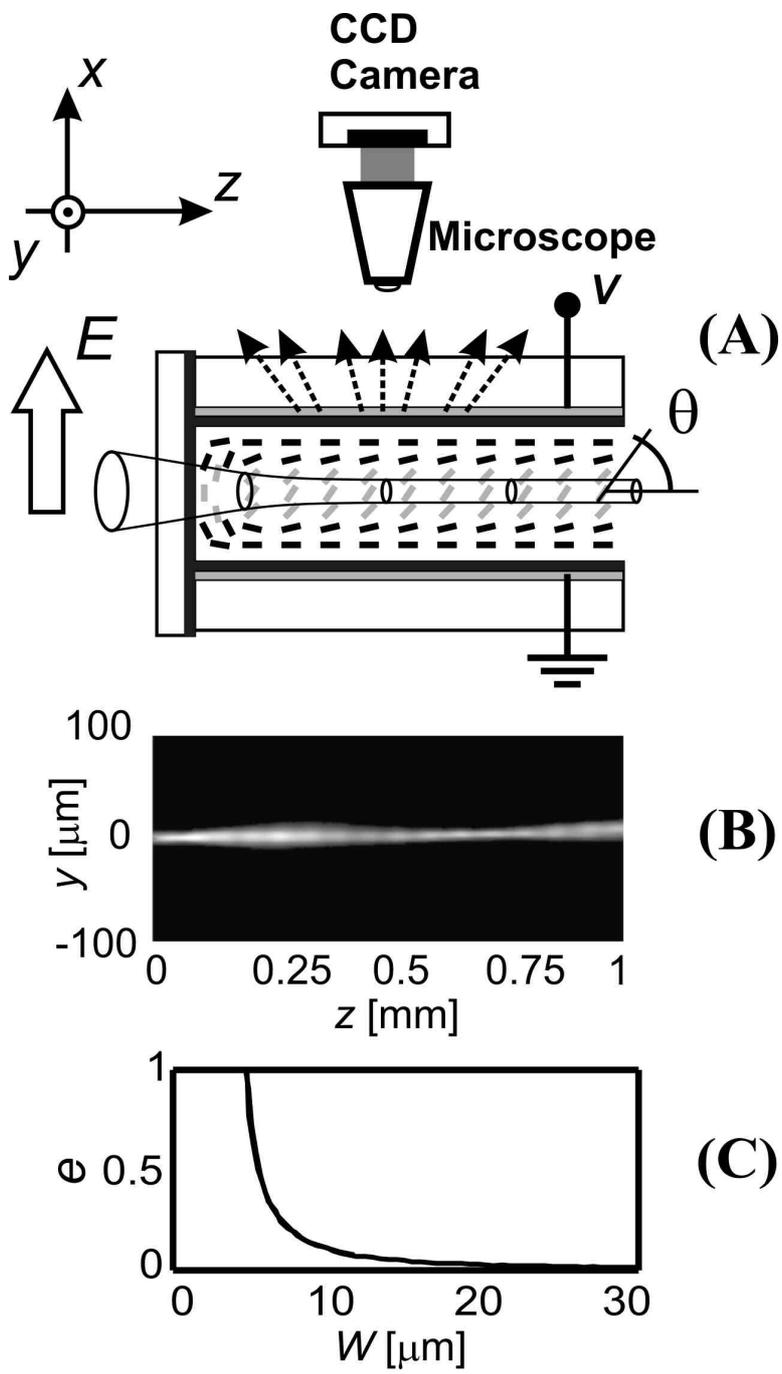

Figure 1

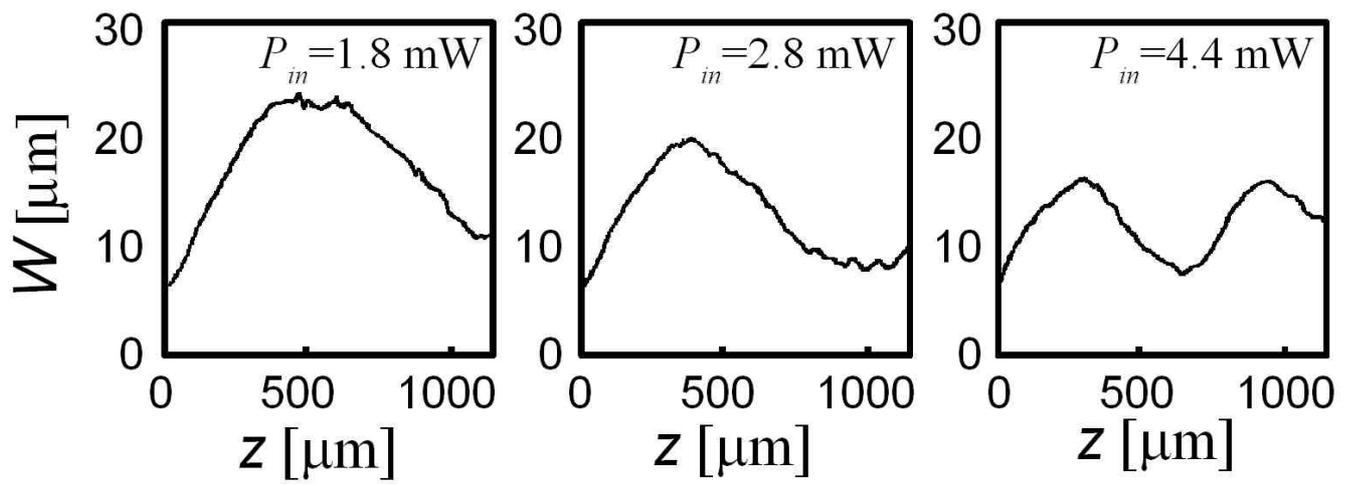

Figure 2

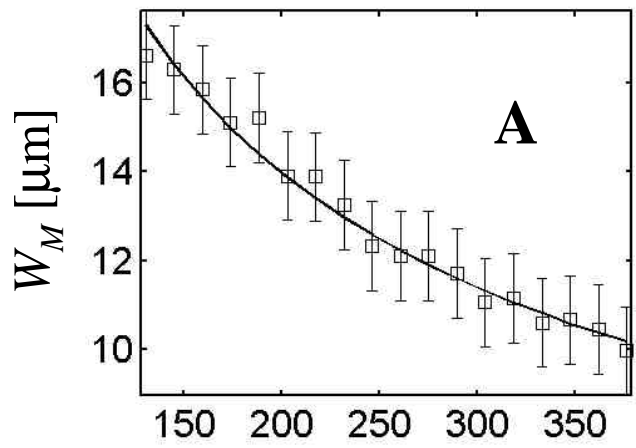
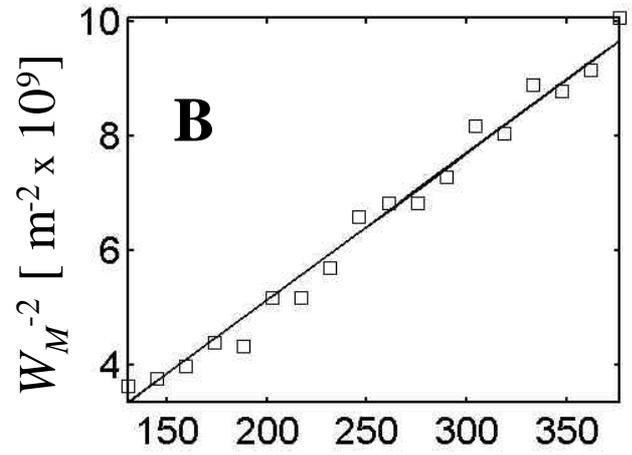
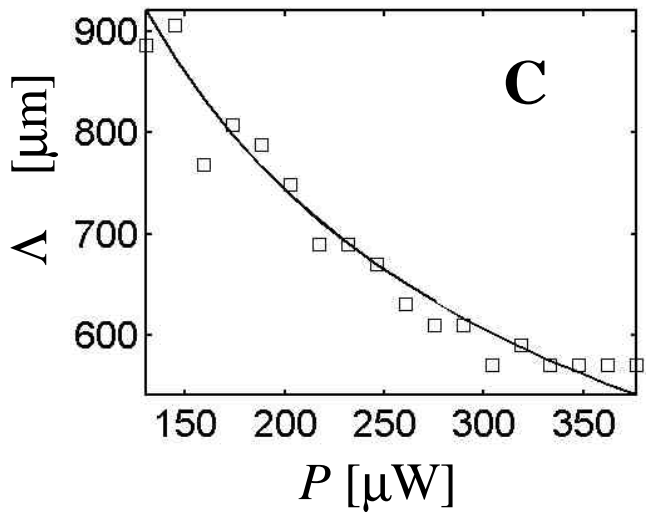
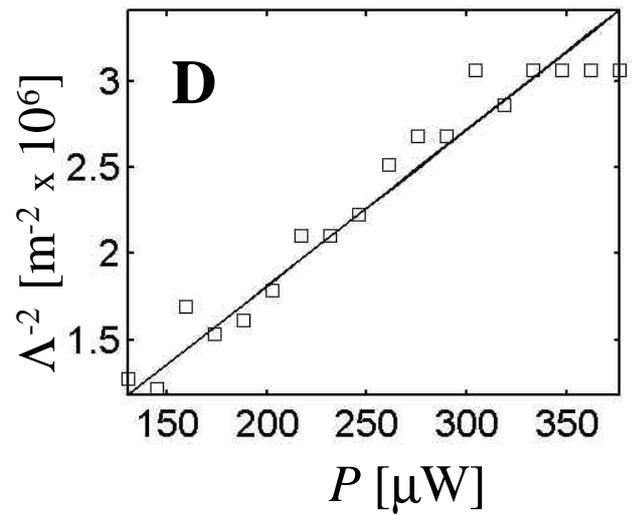

Figure 3